# Quantum computational advantage using photons

Submitted version

Han-Sen Zhong[1,2], Hui Wang[1,2], Yu-Hao Deng[1,2], Ming-Cheng Chen[1,2], Li-Chao Peng[1,2], Yi-Han Luo[1,2], Jian Qin[1,2], Dian Wu[1,2], Xing Ding[1,2], Yi Hu[1,2], Peng Hu[3], Xiao-Yan Yang[3], Wei-Jun Zhang[3], Hao Li[3], Yuxuan Li[4], Xiao Jiang[1,2], Lin Gan[4], Guangwen Yang[4], Lixing You[3], Zhen Wang[3], Li Li[1,2], Nai-Le Liu[1,2], Chao-Yang Lu[1,2, *], Jian-Wei Pan[1,2, *]

Corresponding authors: cylu@ustc.edu.cn (C.-Y.L.) pan@ustc.edu.cn (J.-W.P.)

[1]Hefei National Laboratory for Physical Sciences at Microscale and Department of Modern Physics, University of Science and Technology of China, Hefei, Anhui 230026, China

[2]CAS Centre for Excellence and Synergetic Innovation Centre in Quantum Information and Quantum Physics, University of Science and Technology of China, Shanghai 201315, China

[3]State Key Laboratory of Functional Materials for Informatics, Shanghai Institute of Microsystem and Information Technology, Chinese Academy of Sciences, 865 Changning Road, Shanghai 200050, China

[4]Department of Computer Science and Technology and Beijing National Research Center for Information Science and Technology, Tsinghua University, China

H.-S.Z, W.H, Y.-H.D, and C.-M.C. contributed equally.

**Abstract**

**Gaussian boson sampling exploits squeezed states to provide a highly efficient way to demonstrate quantum computational advantage. We perform experiments with 50 input single-mode squeezed states with high indistinguishability and squeezing parameters, which are fed into a 100-mode ultralow-loss interferometer with full connectivity and random transformation, and sampled using 100 high-efficiency single-photon detectors. The whole optical set-up is phase-locked to maintain a high coherence between the superposition of all photon number states. We observe up to 76 output photon-clicks, which yield an output state space dimension of $\sim 10^{30}$ and a sampling rate that is $\sim 10^{14}$ faster than using the state-of-the-art simulation strategy and supercomputers. The obtained samples are validated against various hypotheses including using thermal states, distinguishable photons, and uniform distribution.**

## Main text

Extended Church-Turing thesis is a foundational tenet in computer science, which states that a probabilistic Turing machine can efficiently simulate any process on a realistic physical device (*1*). In the 1980s, Richard Feynman observed that many-body quantum problems seemed difficult for classical computers due to the exponentially growing size of the quantum state Hilbert space. He proposed that a quantum computer would be a natural solution (*2*).

A number of quantum algorithms have since been devised to efficiently solve problems believed to be classically hard, such as Shor's factoring algorithm (*2*). Building a fault-tolerant quantum computer to run Shor's algorithm, however, still requires long-term efforts. Quantum sampling algorithms (*3-6*), based on plausible computational complexity arguments, were proposed for near-term demonstrations of quantum computational speedup in solving certain well-defined tasks compared to current supercomputers. If the speedup appears overwhelming such that no classical computer can perform the same task in a reasonable amount of time and is unlikely overturned by classical algorithmic or hardware improvements, it was called quantum computational advantage or quantum supremacy (*7, 8*). Here, we use the first term.

A very recent experiment on a quantum processor with 53 superconducting qubits has generated a million noisy (~0.2% fidelity) samples in 200 seconds (*10*), while the first estimation would take about 10,000 years on a state-of-the-art classical supercomputer. It was later argued that the classical algorithm can be improved to cost only a few days to compute *all* the $2^{53}$ quantum probability amplitudes and generate ideal samples (*12*). Therefore, if the task were to output a much larger size of samples, for example, ~$10^{10}$, the computational advantage would be reversed. This sample size dependence of the computational advantage—an analog to the loopholes in Bell tests (*13*)—suggests that a quantum computational advantage would require long-term competitions between faster classical simulations and improved quantum devices. Such efforts are also important for gaining increasingly more evidence against the Extended Church-Turing thesis.

Boson sampling (*5*), proposed by Aaronson and Arkhipov as the first feasible protocol for quantum supremacy, can avoid the sample size loophole. In boson sampling and its variants (*14-17*), non-classical light is injected into a linear optical network, and the output highly random, photon-number- and path-entangled state is measured by single-photon detectors. The dimension of the entangled state grows exponentially with both the number of photons and the modes, which fast renders the storage of the quantum probability amplitudes impossible. The state-of-the-art classical simulation algorithm is to calculate one probability amplitude (Permanent of the submatrix) at a time. The Permanent is classically hard and at least one Permanent should be evaluated for each sample. In addition, boson samplers use photons which can be operated at room temperature and are robust to decoherence, and are appealing to show the surprising yet remarkable computational complexity (*5,19*).

Inspired by the original proposal of Aaronson and Arkhipov (*5*), numerous experiments have demonstrated boson sampling with continuously increasing complexity (*20-27*). However, scaling up boson sampling to a computationally interesting regime remained an outstanding experimental challenge. The early proof-of-principle demonstrations of boson sampling (*20-23*) used probabilistic, post-selected pseudo-single photons from parametric down-conversion (PDC) (*28*). Improved on-demand single-photon sources based on semiconductor quantum dots were developed and employed to significantly increase the multi-photon sampling rates (*25-27*), which culminated at 14-photon detection (*27*). In parallel, new schemes, such as scattershot boson sampling (*14*) and boson sampling with photon loss (*15*), were formulated and demonstrated (*29-31*) to enhance the multi-photon detection rate.

Recently, Gaussian boson sampling (GBS) (*16*, *17*) has emerged as a new paradigm that can not only provide a highly efficient approach to large-scale implementations but also offer potential applications in graph-based problems (*32-35*), point processes (*36*), and quantum chemistry (*37-39*). Instead of using single photons, GBS makes full use of the Gaussian nature of the PDC sources and utilizes single-mode squeezed states (SMSS) as input non-classical light sources which can be deterministically prepared. Sending $k$ SMSSs through an $m$-mode interferometer and sampling output scattering events using threshold detectors (see Fig. 1), it has been shown that the output distribution is related to a matrix function called Torontonian (*17*). Computing the Torontonian, which is an infinite sum of Hafnians, is a computationally hard problem in the complexity class #P. Many efforts have been devoted to benchmarking the classical computational capacity of calculating the Hafnians and Torontonians on supercomputers using continuously optimized algorithms (*40-43*). The most efficient method so far reported that it takes about two days to evaluate a Torontonian function for a 50-photon click pattern (*43*).

Although there were small-scale demonstrations of GBS with up to five photons (*44,45*),

implementing a large-scale GBS faced a number of experimental challenges. First, it requires a large number of SMSS sources with sufficiently high squeezing parameters, photon indistinguishability, and collection efficiency, simultaneously (*46*). Second, all the photons are interfered inside a sufficiently large and deep interferometer with full connectivity, matrix randomness, near-perfect wave-packet overlap and phase stability, and near-unity transmission rate, simultaneously. Third, as the GBS replies on coherent superposition of photon number, phase control of all the SMSSs is required from their generation to propagation inside the interferometer. Fourth, high-efficiency detectors are needed to sample the output distribution. Finally, the obtained sparse samples from a huge output Hilbert space should be validated, and the performance of the GBS should be benchmarked and compared with a supercomputer.

We start by describing the design of the quantum light source arrays. We use a custom-designed laser system consisting of a Mira 900, a pulse shaper, and a RegA 9000, which outputs an average power of 1.4 W at a repetition rate of 250 kHz, a central wavelength of 776 nm, and a pulse width of 200 fs (see Fig. S1). We choose a repetition rate much lower than commonly used ~80 MHz femtosecond lasers for two reasons: increasing single-pulse pump energy by a factor of ~300 to generate high squeezing parameter, and relaxing recovery time to ~μs for single-photon detectors to eliminate readout error. We use a deformable mirror to cancel unwanted chirp and dispersion of the laser pulse, reshaping it to reach Fourier transform limit (see Fig. S2 for its spectral intensity and phase), which is necessary to ensure high indistinguishability of the PDC photons.

The laser is split into 13 paths with equal intensity, and focused on 25 individual PPKTP crystals (see Fig. 2A and Fig. S3, S4) with a waist of ~60 μm to produce 25 two-mode squeezed states (TMSSs) as input, which, due to our hybrid-encoding (which will be discussed later) unitary transformation, is equivalent to 50 SMSSs. The relative phase and squeezing parameter for each pair are shown in Fig. 2B. To obtain a high photon purity, the photons should be free from any undesired correlation in their spatiotemporal degrees of freedom, that is, the two-photon joint amplitude function is factorable. The PPKTP crystals are designed (Fig. S3) and temperature-controlled (Fig. S5) to generate degenerate and frequency-uncorrelated photon pairs, as confirmed by joint spectrum in Fig. 2C which predicts a spectral purity of 0.98. The purity is increased to 0.99 using a 12-nm filtering (Fig. S6, S7). A second estimation of the purity is by unheralded second-order correlation measurements (*47*), which agrees with independent calibrations using time-delayed Hong-Ou-Mandel (*48*) interference visibility at low pump power. The measured pair-wise photon purities are plotted in Fig. 2D, which have an average of 0.938. The decrease of the photon indistinguishability compared to the prediction from the joint spectra is mainly caused by self-phase modulation. As shown in Fig. 2E, the measured average collection efficiency is 0.628, which could be further improved using a larger focal spot (*49*).

In contrast to Aaronson-Arkhipov boson sampling (*5*) where there is no phase relation between single photons, GBS relies on a coherent superposition of the photon numbers. Thus, the whole optical setup—from the 25 PPKTPs to the 100-mode interferometer—must be locked to a fixed phase in the presence of various environmental perturbations. To this aim, we develop an active phase locking over the whole optical path and passive stabilization inside the interferometer. For the active locking, as shown in Fig. 3A, a pump laser beam is used as a reference for all the squeezed states. We take advantage of the collinearity of the PDC, that is, the pump laser shares the same propagation path as the TMSSs—both in free space and optical fibers—before entering the interferometer. After propagating through a ~2-m free space and 20-m optical fiber, a ~10 μW laser pulse is separated by a dichromatic mirror, which is then combined on a beam splitter with the reference laser pulse. A balanced detection scheme, which is insensitive to laser power fluctuation, is used to read out the phase information (see SOM).

To overcome the path length fluctuates, we wind 5-meter-length optical fiber around a piezo-electric cylinder which has a sensitivity of 1.5 rad/V, a resonance frequency of 18.3 kHz, and a dynamical range of 300 rad (see Fig. 3A and SOM). The optical fibers are temperature stabilized to minimize the slow drift. A typical time trace of the observed phase stability is shown in the upper panel of Fig. 3B, which locks the phase of the 776 nm laser with a standard deviation of 0.04 rad (~5 nm, see SOM). After the dichromatic mirror—the ending point of the active phase locking, the squeezed states are then sent into the multi-mode interferometer, as shown in Fig. 3C. We apply passive phase stabilization to the interferometer by adhering the devices onto an ultra-low-expansion glass plate which is temperature stabilized within 0.02ºC. The drift is $\lambda/180$ in 3.5 hours (see bottom panel of Fig. 3B). For the whole system, as shown in Fig. 3D, the high-frequency noise standard deviation is $\lambda/350$ and the low-frequency drift is $\lambda/63$ within 1 hour, a time sufficient for completing the sampling and characterizations. We estimate that the photon interference visibilities drop are less than 1% due to the phase instability.

We make use of both the photons' spatial and polarization degree of freedom to realize a 100×100 unitary transformation (*20, 50*). Here, the mode mapping is $\{1, 2, \cdots 100\} = \{|H\rangle_1 |V\rangle_1 |H\rangle_2 |V\rangle_2 ... |H\rangle_{50} |V\rangle_{50}\}$, where *H* (*V*) denotes horizontal (vertical) polarization, and the subscripts denote the spatial mode in the interferometer. We adopt a compact 3D design for the 50-spatial-mode interferometer, which simultaneously combines near-perfect phase stability and wave-packet overlap, full connectivity, random matrix, and near-unity transmission rate (Fig. 3C and SOM). This optical network effectively consists of 300 beam splitters and 75 mirrors (see Fig. S9). A much more stringent requirement on the fabrication precision is required, compared to the previous design (*27*), because the coherence length of the PDC photons is about two orders of magnitude shorter than those of the quantum-dot single photons. The relative path delay between different ports is measured with a standard deviation of 2.6 μm (see SOM), which is much smaller than the photons' coherence length (~88 μm) and causes a 0.2% drop of the interference visibility between independent photons. The transmission rate of the interferometer is measured to be 97.7%, and the average coupling efficiency in all the output ports is ~90%.

In the output of this interferometer, half- and quarter-wave plates set at random angles are used to apply random transformations between the two orthogonal polarizations, which are then split into two spatial modes using 50 polarizing beam splitters. See Fig. S13 for a photograph for the interferometric part of our experimental set-up. Finally, the 100 output modes are detected by 100 superconducting nanowire single-photon detectors with an average efficiency of 81% (see SOM).

Contrary to the Aaronson-Arkhipov boson sampling where the sampling matrix is given solely by the interferometer, the GBS matrix absorbs both the unitary transformation of the interferometer and the squeezing parameters and phases of the Gaussian input state. The squeezing parameters for each TMSS can be calculated individually by measuring their photon number distribution. The relative phases between the 25 different TMSSs are measured using small-scale GBS tests and fitting (see SOM). The unitary matrix for the 3D interferometer is measured using a narrow-band laser using the method as in (*27*). With these parameters together, we reconstruct the corresponding unitary matrix of the spatial-polarization hybrid encoded 100×100 interferometer as plotted in Fig. 3E and 3F for the elements of amplitudes and phases, respectively. Figure S14 shows that the product with its Hermitian conjugate gives an identity matrix, which confirms that the obtained matrix is unitary. For the hardness arguments of the boson sampling to hold, the matrix should be randomly drawn according to the Haar measure. Thus, the measured elements as shown in Fig. 3E and Fig. 3F are compared with the ideal Haar-random matrix elements. The statistical frequency of the measured 5,000 elements of amplitude and phase is shown in Fig. S15, which agrees with the prediction from Haar-random matrix.

We can divide our experiment into three different regimes which we call as easy regime, sparse regime, and intractable regime. The easy regime is at small scales where we can obtain the full output distribution, as in most previous proof-of-principle experiments. The size of the output Hilbert spaces is a function of both the photon and mode number, which grows exponentially. When the photon click number is well above 4, then only a small fraction of output combinations can obtain one event, while most output will have zero events, which is called the sparse regime. There it becomes infeasible to register the whole output distribution, but only sampling is possible. An important yet open question is how to verify the large-scale GBS. Unlike Shor's algorithm (*3*) where its solution can be efficiently verified; for the GBS, a full certification of the outcome is strongly conjectured to be intractable for classical computation (*5*). There have been various validation methods (*51-54*) proposed and related experiments (*24-27,29-31,44, 45,55-58*) to gather supporting or circumstantial evidence for the correct operations of the boson-samplers. However, when the output click number exceeds ~30, even the calculations required for the validation methods become intractable for classical computers (*43*).

We start describing our experimental results from the easy regime. We test with three pairs of input TMSSs and the output photon click number is 2. The output distribution is plotted in Fig. 4A. We use fidelity (*F*) and total variance distance (*D*) to characterize the obtained distribution, defined by: $F = \sum_i \sqrt{p_i q_i}$, and $D = \sum_i |p_i - q_i|/2$ ($p_i$ and $q_i$ denote the theoretical and experimental probability of *i*-th basis, respectively). For a perfect boson-sampler, the fidelity should equal to 1 and the distance should be 0. The measured average fidelities and distances are 0.990(1) and 0.103(1). The fidelities and distances for all the 23 different input configurations with three TMSSs input are shown in Fig. 4B, which confirms that the GBS works properly.

Now we move to the sparse and intractable regime. Using 25 TMSSs input, the output photon number distribution using threshold detectors is plotted in Fig. 4C. The average click number is 43. Within 200 s, we obtain 3,097,810 43-photon coincidence events, and one 76-photon coincidence. The state-space dimension of our experiment is plotted in Fig. 4D, reaching up to $10^{30}$, which is 14 and 16 orders of magnitude larger than the previous experiments using superconducting qubits (*10*) and single photons (*27*).

Although we cannot directly calculate and verify the results in the large photon number regime, we hope to provide strong evidence that the large-scale GBS device continues to be governed by quantum mechanics when it reaches quantum advantage regime. The credibility of the certification processes relies on gathering circumstantial evidence while ruling out possible hypotheses plausibly to occur in this experiment. We validate the desired input TMSSs against input photons that are thermal states—which would result from excessive photon loss—and distinguishable—which would be caused by mode mismatch. First, we compare the obtained output click number distribution with results from classical simulations of the GBS using thermal light and distinguishable SMSSs (Fig. 4E). Evidently, there are strong deviations in their line shapes and peak positions, which support that the obtain distribution indeed arises from genuine multi-photon quantum interference.

We further investigate two-point correlation (*59*), which is derived from the Hanbury-Brown-Twiss experiment, to reveal the non-classical properties of the output light field. Here, the two-point correlation between the mode *i* and the mode *j* is defined as: $C_{i,j} = \langle \Pi_1^i \Pi_1^j \rangle - \langle \Pi_1^i \rangle \langle \Pi_1^j \rangle$, where $\Pi_1^i = \mathbf{I} - |0\rangle_i \langle 0|_i$ represent a click in mode *i*. We calculate the distribution of all $C_{i,j}$ for the experimentally obtained samples, which is then compared with those from theoretical predictions, thermal states hypothesis and distinguishable SMSSs hypothesis. As shown in Fig. 4F, the statistic of experimental samples significantly diverge from the two hypotheses and are in a good agreement with the theoretical prediction.

Having studied the whole distribution, second, we closely look into each subspace with a specific photon click number. We develop a method called heavy output generation (HOG) ratio test (see SOM). Figure 4G shows typical examples of HOG analysis for photon clicks from 26 to 30, which show a stark difference between TMSS with thermal states. We emphasize that, as the output average photon number is ~43, the tested 26-30 click regime—that shares the same set-up as higher photon number—is actually in the post-selected subspace that effectively suffers from more photon loss than in the regime with a larger number of clicks. Therefore, although we can only verify the 26-30 click regime, the events with larger number of clicks in the output—the intractable regime—can be validated against the thermal state hypothesis with higher confidence.

We continue to rule out another important hypothesis that boson sampling output would be operationally indistinguishable from a uniform random outcome, one of the earliest criticize (*51*) to boson sampling. In stark contrast, due to constructive and destructive interference, an ideal boson-sampler is expected to generate samples with exponentially different occurring probability (*5,51*). We develop a method (see SOM) to reconstruct the theoretical probability distribution curve shown in Fig. 4H. We can match each obtained sample to the theoretical curve, as illustrated by the blue data points and vertical blue lines in Fig. 4H (see Fig. S24 for more data). The density—the frequency of occurrence—of the blue lines is in a good agreement with the theoretically calculated probability, which intuitively indicates that our results cannot be reproduced by a uniform sampler.

Finally, we estimate the classical computational cost to simulate our GBS machine. We have benchmarked GBS on the Sunway TaihuLight using a highly optimized algorithm (*43*). The time cost to calculate one Torontonian scales exponentially as a function of output photon clicks, from ~0.03 s for 30 photons to ~2 days for 50 photons in 256 bit precision. Moreover, to obtain one sample using Markov Chain Monte Carlo sampling method (*18*), one usually needs to calculate ~100 Torontonians of candidate samples. As shown in Fig. 4C, the GBS simultaneously generates samples of different photon-number coincidences, which can be seen as an intrinsically high-throughput sampling machine. For each output channel and the registered counts in Fig. 4C, we calculate the corresponding time cost for the supercomputer, which is plotted in Fig. 5. The main classical computational time cost is from photon number above 50 and peaks at 70. The line shape is due to a competition between the increased computational complexity and reduced counts at higher number of photons. Summing over the data points in Fig. 5, we estimate that the required time cost for the TaihuLight (Fugaku) to generate the same amount of samples in 200 s from the GBS device would be $8\times10^{16}$ s ($2\times10^{16}$ s), which is 2.5 (0.6) billion years.

We hope our work will inspire new theoretical efforts for quantitative characterizations (*60*) for large-scale GBS, and to improve the classical simulation strategies (*17*, *40-43*, *61*, *62*) optimized for our experimental parameters, to challenge the observed quantum computational advantage on the order of $10^{14}$. For example, there migh be alternative classical algorithms which are expected to be much faster for low rank matrices than the present used method, but they become slower for nearly full rank matrices. As the maximum rank ratio in our experiment is ~50:75, our method keeps optimal over the alternatives. A lower cost algorithm for non-full rank matrices could be achieved in the future, but currently there is no known algorithm to take advantage of this. Furthermore, classical computational speedup in generating noisy samples out of a GBS with photon loss is another interesting direction.

In the near term, our GBS machine can be further upgraded using higher-efficiency photon detectors, large-scale interferometer, and stimulated PDC with lower pumping power and higher collection efficiency (*63*), in order to obtain increasing experimental evidence against the extended Church-Turing thesis (*1*).

In addition to the sampling-based quantum computational advantage, the GBS links to several potentially practical applications including certifiable random numbers, graph optimization (*32-34*), graph similarity (*35*), point process (*36*), molecular docking (*37*), quantum chemistry (*38,39*), and quantum machine learning (*64*). A natural next step would be to use the GBS quantum computer developed in our experiment as a special-purpose photonic platform to investigate these real-world applications, as a step toward noisy intermediate-scale quantum processing (*65*). Finally, the quantum optical set-up consisting of squeezed states fed into a linear optical network and followed by post-selection measurement outcome, can be used as a factory for heralded production of different family of quantum states, such as Fock state, cat state, and NOON state (*66*).

*Note added:* A new algorithm reported a quadratic speedup in simulating non-collision GBS [arXiv.2010.15595]. We thank the authors for helpful private communications. In our work, collision modes dominate. The actual number of photons is approximately twice the number of clicked detectors, and thus the new algorithm cannot be readily used to speed up simulating our current experiment.

*Note added 2*: In the revised version, validations against thermal states and indistinguishable photons have been performed up to 40 clicks.

*Note added 3*: An updated experiment with considerably improved SMSS collection efficiency and higher mode number (144) is being performed at its final stage at the time of the release of this paper.

The supplemental materials which is 80+ pages and 23+ MB is made available to download at http://staff.ustc.edu.cn/~cylu/lightquantadvantage.pdf.

**Figure Captions**

**Fig. 1 | An overview of our GBS implementation.** The GBS device is realized by sending *50* single-mode squeezed states (SMSSs) into a 100-mode interferometer, and then sampling from the output distribution using 100 single-photon detectors.

**Fig. 2 | Quantum light sources for the GBS.** (A) An illustration of the experimental set-up for generating TMSSs. A spectrally and spatially shaped pulse laser (Fig. S1, S2) is split into 13 paths (Fig. S3, S4) and then focused onto 25 PPKTP crystals to generate down-converted photons in squeezed vacuum state. Every crystal is placed on a thermoelectric cooler (TEC) for central-wavelength-degeneracy tuning. The down-converted photons are isolated from the pumping laser by a dichromic mirror (DM), the time walk between *H*- and *V*-polarized photons are compensated by a KTP crystal, before being collected into a single-mode-fiber. (B) Wigner function of all 25 sources. The squeezing parameter *r* and phase $\phi$ of each source is presented as $(r, \phi)$ in each panel. (C) The measured joint spectrum of the photon pairs, which indicates that the two photons are free of frequency correlations. (D) Purity diagram of the 25 sources. The measured average purity is 0.938, obtained by unheralded second-order

correlation measurement. (**E**) Diagram of the efficiency of the 25 sources. After a 12-nm filter, the measured average efficiency is 0.628.

**Fig. 3 | Phase-locking from the photon sources to the interferometer.** (**A**) Schematic diagram of the active phase-locking system. Every pumping laser sharing the same propagation path with the corresponding TMSS is sent into a PBS to interference with a reference laser beam. The phase is extracted from the detected heterodyne interference signal. To lock the phase to the reference beam in real time, the feedback control signal given by PID controller is sent to a piezo-electric cylinder winded by 5-m fiber to frequently compensate phase fluctuations. (**B**) Phase stability tests in experimental system. The upper (bottom) panel is a typical monitoring of phase fluctuation of active (passive) phase locking over 3.5 hours. The measured standard deviation of the phase is as small as 0.02 rad ($\lambda/150$) (0.017 rad ($\lambda/180$)). (**C**) Experimental set-up of 100-mode photonic network adhered on a 200×200 mm ultra-low-expansion plate. The blue light path is employed to make all 25 pumping lasers interference with 25 reference lasers. The red part is arranged to let 25 TMSSs pass through photonic network and scattered into 100 spatial modes. (**D**) A typical phase stability measurement of the whole system over 50 minutes. (**E**) The diagram of measured 5000 amplitude elements. (**F**) The diagram of measured 5000 elements of matrix phase.

**Fig. 4 | Experimental validation of GBS.** (**A**) Experimental (red line) and theoretical (blue line) two-photon distribution with three TMSSs input. (**B**) Summary of statistical fidelity and total variation distance of two-photon distribution for 23 different input sets. The average fidelity and total variation distance are 0.990(1) and 0.103(1), respectively. (**C**) The output photon number distribution with all 25 TMSSs in. The average detected photon number is 43, while the maximal detected photon number is up to 76. (**D**) Summary of dimensions of output state space. (**E**) The photon number distributions of the experiment (red), thermal state (blue) and distinguishable SMSS (green), respectively. The deviations of the line shape and peak positions indicate that our experiment is far from these two hypotheses. (**F**) Two-photon correlation statistics for all 2-mode combinations. The statistic of the experimental results (red) highly overlap with the theoretical predictions (orange), and deviate significantly from the thermal state hypothesis (blue) and the distinguishable SMSS hypothesis (green). (**G**) Heavy out generation (HOG) ratio test to further exclude thermal-state hypothesis with detected photon number ranging from 26 to 30. All HOG ratios of thermal state quickly converge to 0, which confirm that our experiments are from genuine GBS. (**H**) Heavy output generation of samples to exclude uniform distribution. The gray line is the calculated theoretical distribution by uniformly sampling the output space. The experimental data (blue dot) is mapped to the theoretical distribution, which clearly demonstrates the heavy output generation of experimental samples, thereby directly exclude the uniform distribution.

**Fig. 5 | Classical computational cost.** The estimated time cost on Sunway TaihuLight supercomputer. The error bar is calculated from Poissonian counting statistics of the raw detected events.

**Acknowledge:** This work is dedicated to the people in the fight against the COVID-19 outbreak during which the final stage of this experiment was carried out. We thank Jelmer Renema, J. P. Dowling, Christian Weedbrook, Nicolas Quesada, Yi Jiang, Jin-Wei Jiang, Si-Qiu Gong, Bai-Bo Wang, Yu-Huai Li, Hao-Wen Cheng, Qi Shen, Yuan Cao, Yaojian Chen, Haitian Lu, Haohuan Fu, and Teng-Yun Chen for very helpful discussions and assistance. **Funding:** This work was supported by the National Natural Science Foundation of China, the National Key R&D Program of China, the Chinese Academy of Sciences, the Anhui Initiative in Quantum Information Technologies, and the Science and Technology Commission of Shanghai Municipality. **Author contributions:** C.-Y.L. and J.-W.P. designed and supervised the research. H.-S.Z., M.-C.C, and J.Q. developed the theory. H.-S.Z., H.W., Y.-H.D., L.-C.P., Y.-H.L., J.Q., D.W., X.D., L.-L., N.-L.L. and C.-Y.L. carried out the optical experiment and collect the data. Y.H. and X.J. designed the 100-channel counter. C.-M.C., Y.L, P.H., L.G., and G.Y. performed data analysis and validation on a supercomputer. P.H., X.-Y.Y., W.-J.Z, H.L., L.Y., and Z.W. developed single-photon detectors. H.-S.Z., M.-C.C, C.-Y.L. and J.-W.P. analyzed the data and prepared the manuscript. All authors discussed the results, and reviewed the manuscript. **Competing interests:** The authors declare no competing interests. **Data and materials availability:** All data are available in the manuscript or the supplementary materials. Requests for materials should be addressed to C.-Y.L. or J.-W.P.

Fig 1

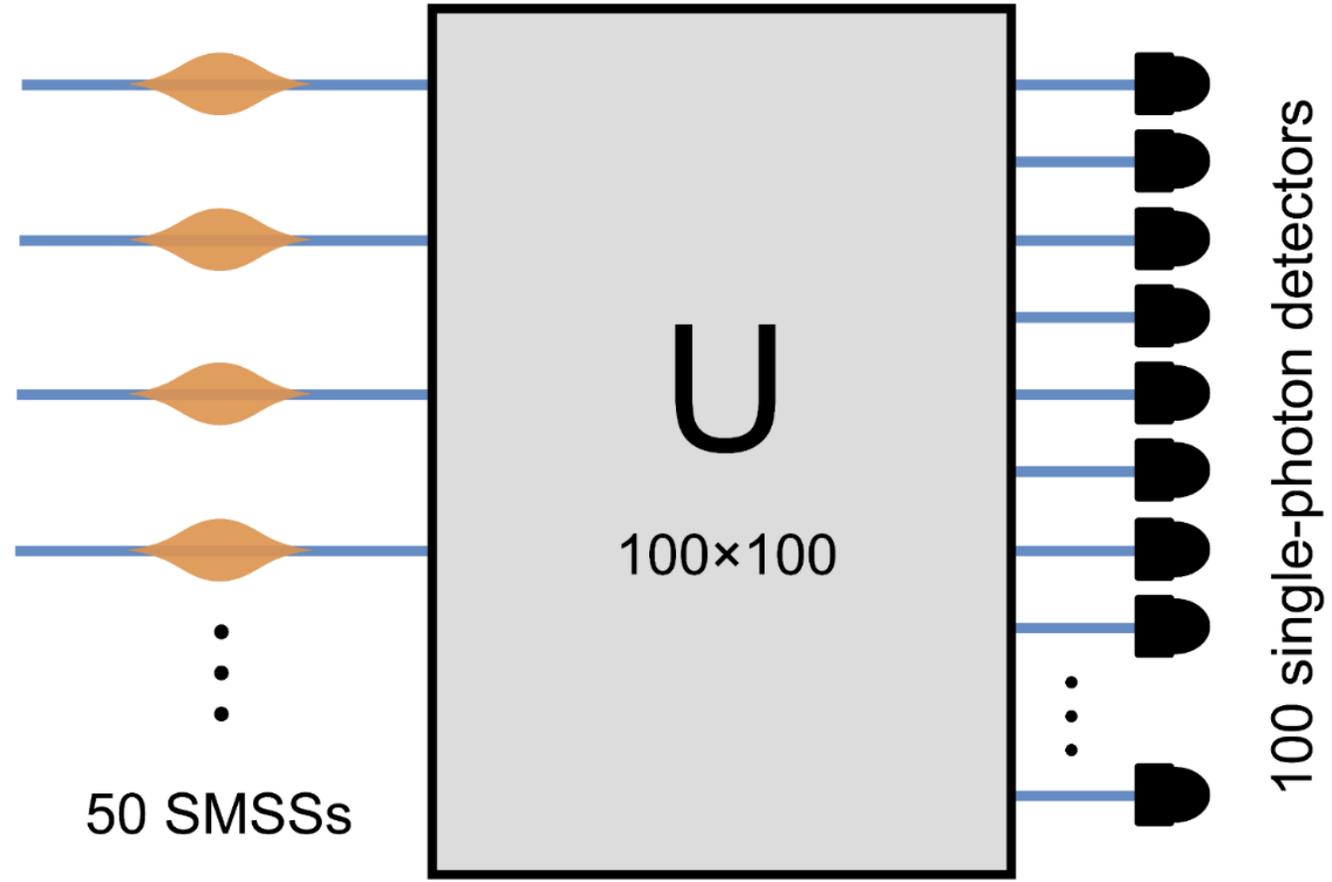

Fig 2

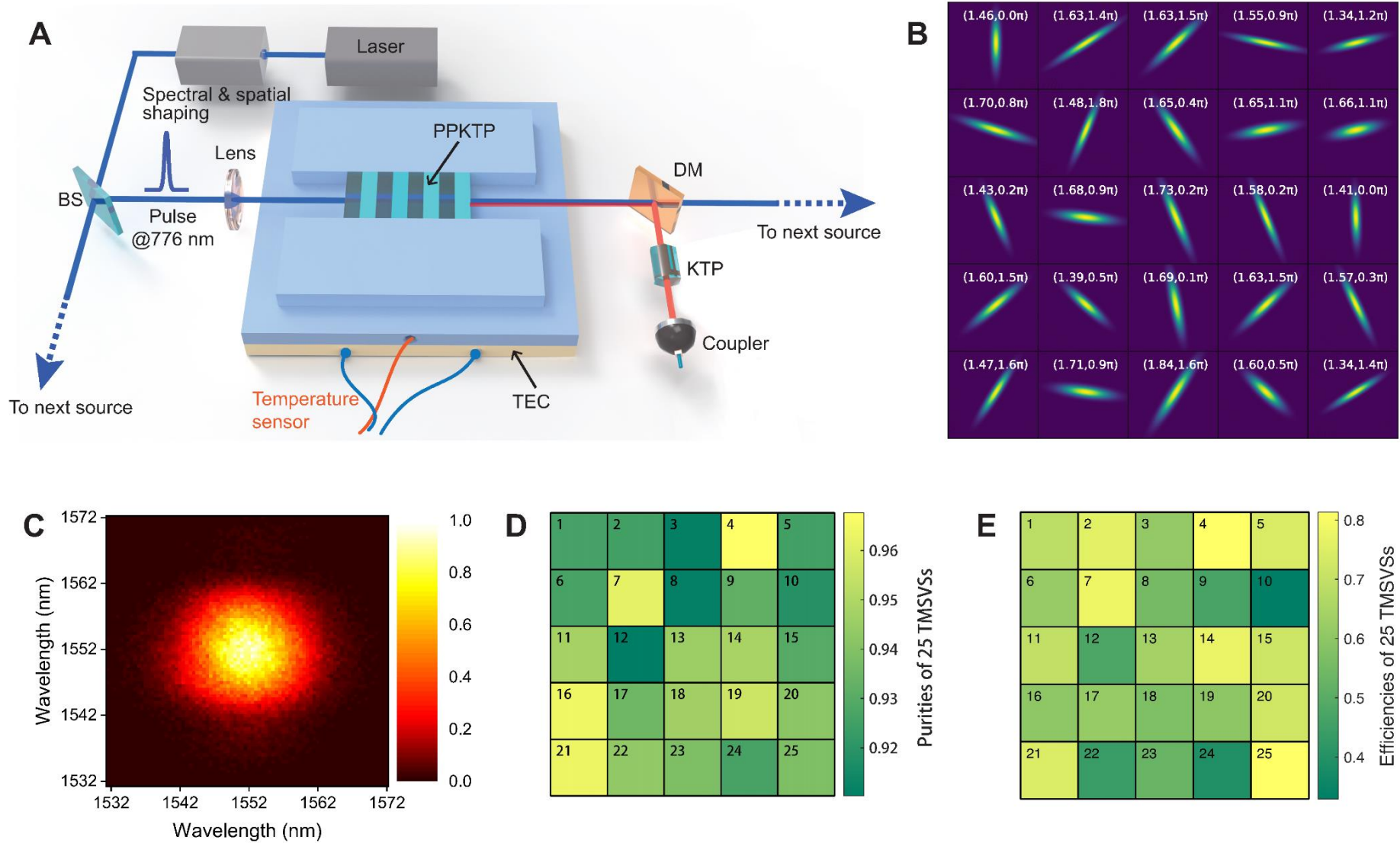

**Fig 3**

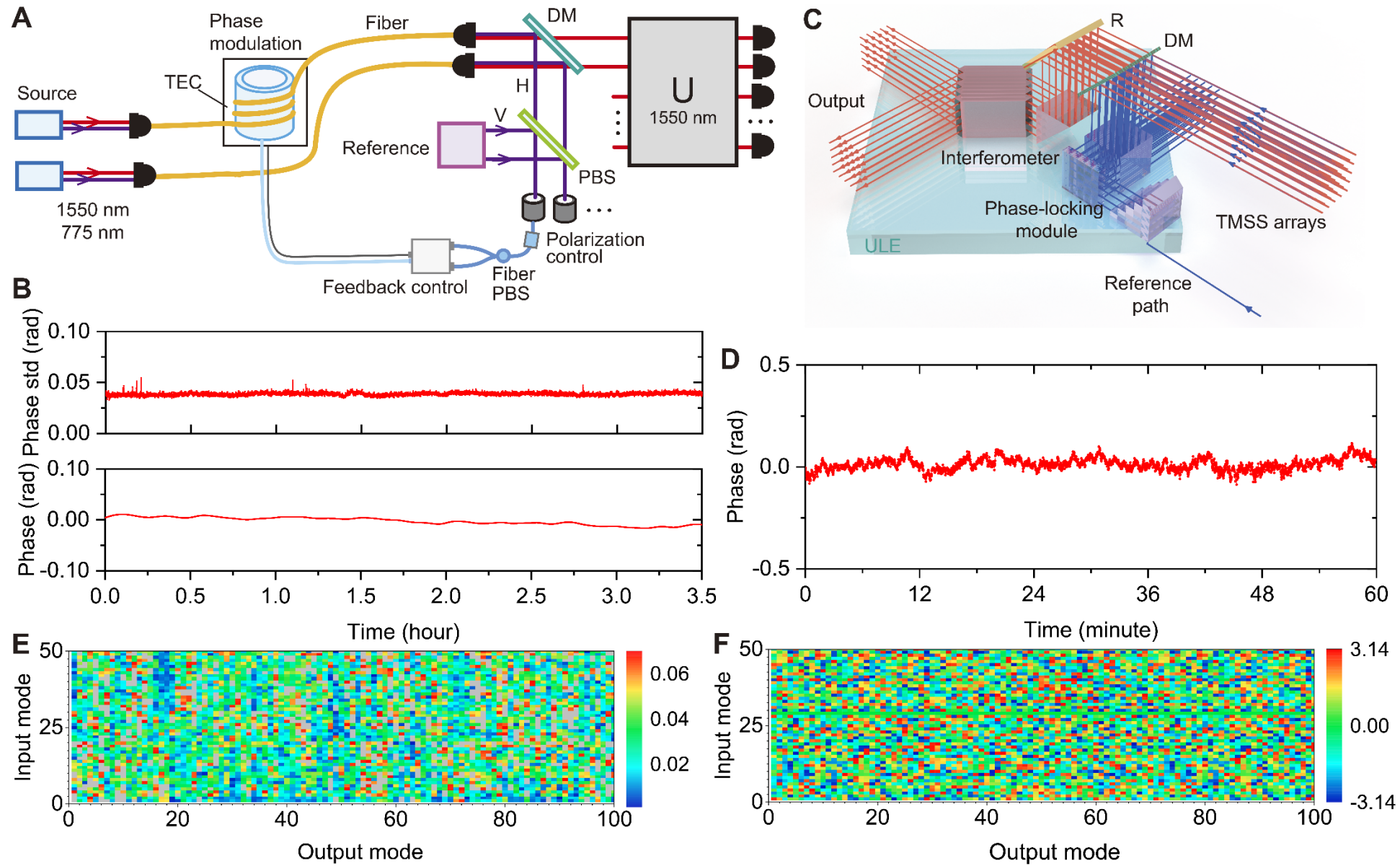

Fig 4

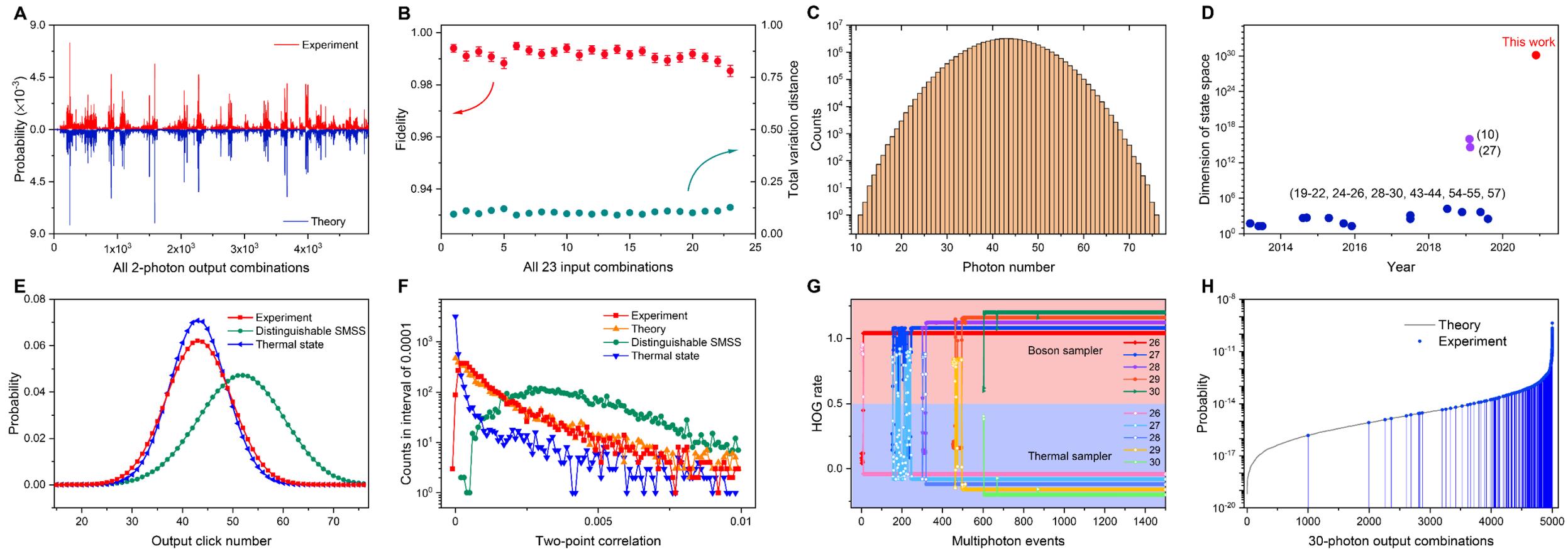

Fig 5

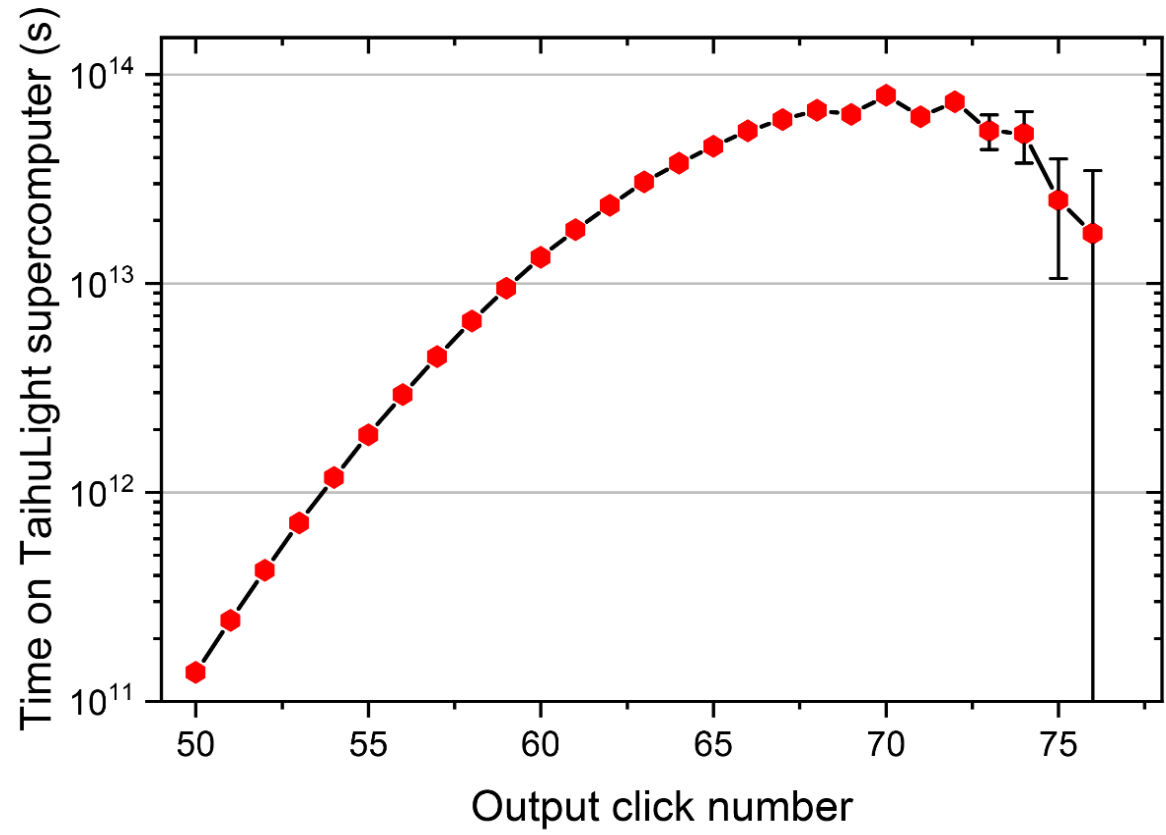

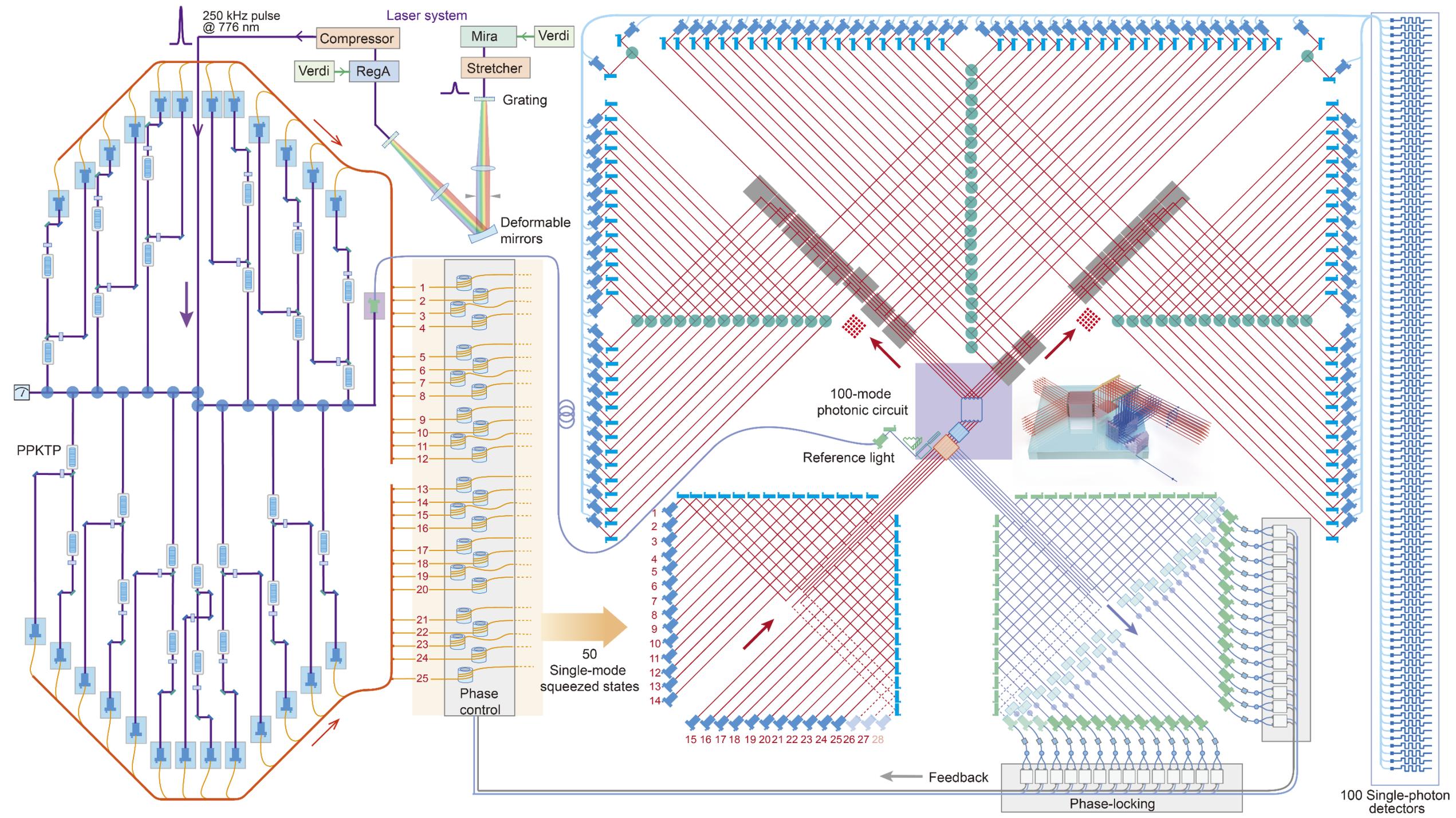

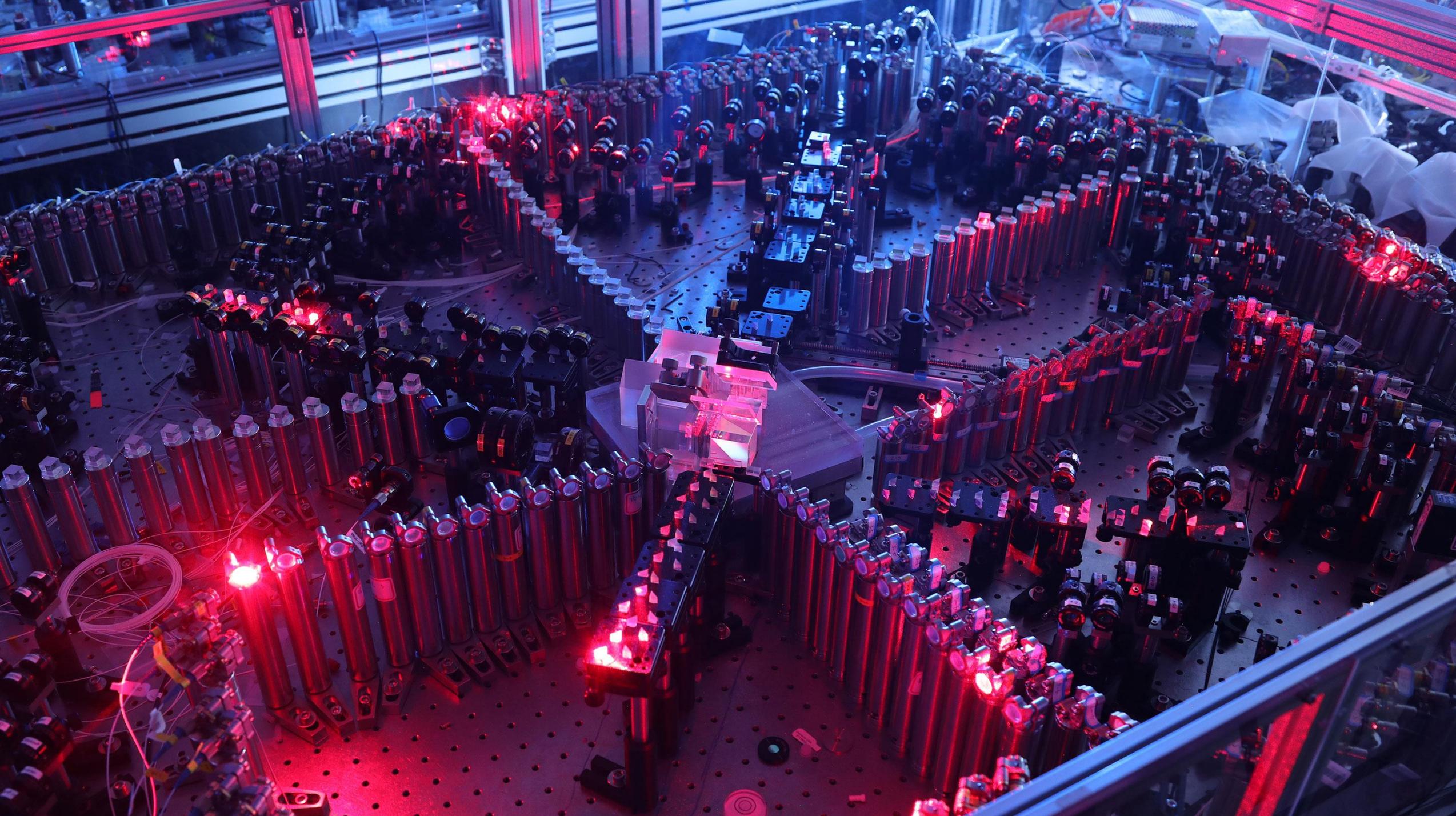